\documentclass{jltp}
\def\lsi{\raise0.3ex\hbox{$<$\kern-0.75em\raise-1.1ex\hbox{$\sim$}}}
\def\gsi{\raise0.3ex\hbox{$>$\kern-0.75em\raise-1.1ex\hbox{$\sim$}}}
\newcommand{\lsim}{\mathop{\lsi}}
\newcommand{\gsim}{\mathop{\gsi}}

\usepackage{graphicx} 

\title{Local Gauge Invariance\\ and Formation of Topological 
Defects\footnote{Talk
given at the ULTI symposium, Pohja, Finland, 10--14 January 2001.}}

\author{Arttu Rajantie}

\address{DAMTP, CMS, Wilberforce Road, Cambridge CB3 0WA, United
Kingdom}

\begin{document}

\maketitle

\markright{\hfill{\rm DAMTP-2001-17}}
\thispagestyle{myheadings}

\runninghead{A. Rajantie}{Formation of Topological Defects}

\begin{abstract}
In superconductors, and in
other systems with a local U(1) 
gauge invariance, there are two mechanisms that form topological
defects in phase transitions. 
In addition to the standard Kibble mechanism,
thermal fluctuations of the magnetic field 
also lead to defect formation.
This mechanism is specific to local gauge theories,
predicts a qualitatively
different spatial defect distribution and is the dominant source for
topological defects
in slow transitions. I review the arguments that lead to
these conclusions and discuss possibilities of testing the scenario in
superconductor experiments.

\end{abstract}

\section{INTRODUCTION}
The Kibble mechanism\cite{Kibble:1976sj} 
gives a very simple and elegant description for
the formation of topological defects in symmetry breaking phase
transitions, and has been
studied intensively in experiments with
liquid crystals\cite{Chuang} and 
liquid Helium.\cite{bauerle,Ruutu:1996qz,ref:He4} 
Even though the results do not
agree completely with the theoretical predictions, the mechanism
itself
is still believed to be correct. Instead, the discrepancies are
presumably due to the extra assumptions one has to make about the
dynamical time and length 
scales of the system in order to derive quantitative
predictions.\cite{Zurek:1996sj,Rivers:2000mb}

More recently, defect formation has also been studied in
superconduc\-tors\cite{ref:carmi} 
and Josephson junctions.\cite{ref:carmi2} 
These systems differ from liquid Helium in a significant way: the
order parameter, i.e., the Cooper pair, is electrically charged, and
the symmetry that is broken in the phase transition is the local U(1)
gauge invariance associated with electromagnetism. This means that the
arguments behind the Kibble mechanism 
cannot be applied directly, and there has been
some discussion over the years on how it should be 
modified.\cite{Kibble:1995aa}

However, it was recently pointed out by myself and 
Hindmarsh\cite{Hindmarsh:2000kd} 
that it is not enough to
simply modify the Kibble mechanism, because there is another, totally
separate effect that leads to defect formation. While the Kibble
mechanism is based on the dynamics of the order parameter field, this
new scenario relies on thermal fluctuations of the magnetic
field. It also leads to predictions that are qualitatively different
from those of the Kibble scenario.

Unfortunately, the experimental setup in Ref.~\onlinecite{ref:carmi} 
was such that
the contribution from our mechanism was negligible. Basically, one
would have to be able to detect the positions and signs of individual
vortices in order to be able to test the mechanism. 

The aim of this paper is to review the mechanism of
Ref.~\onlinecite{Hindmarsh:2000kd} and
to discuss its predictions for a superconducting film. 

\section{KIBBLE MECHANISM}
Let us first review the Kibble mechanism and 
consider a simple two-dimensional toy model with an U(1)
symmetry, i.e., our order parameter $\psi$ is a complex scalar field. 
The thermodynamics of the model is described by the free
energy,\cite{Kleinert} which we approximate in the standard way by
\begin{equation}
\label{equ:Fglobal}
F[\psi]=\int d^2x\left[ \vec{\nabla}\psi^*\cdot \vec{\nabla}\psi
+m^2\psi^*\psi+\lambda\left(\psi^*\psi\right)^2
\right],
\end{equation}
where $m^2\sim T-T_c$.

When the critical temperature is approached from the symmetric
phase, the correlation length of the order parameter, i.e.,
the coherence length $\xi$, diverges but it
cannot grow arbitrarily fast, because at least it is constrained by
the speed of light. Therefore, if the phase transition takes place in
a finite time, the coherence length reaches a finite maximal value
$\hat{\xi}$, and if we divide the system in domains of radius
$\hat{\xi}$, the phase angle of the order parameter is uncorrelated between
different domains. At points where
three or more domains meet, there is a fixed probability that
the phase angle cannot be smoothly interpolated between the domains,
in which case a vortex is formed. Thus,
the number density of vortices must behave as
\begin{equation}
\label{equ:kibblepred}
n_{\rm Kibble}\sim \hat{\xi}^{-2}.
\end{equation}

Because the mechanism is based on very general assumptions only, this
result applies to all systems in which a global U(1) symmetry breaks down 
spontaneously, for instance to $^4$He. It can also be generalized to
other symmetry groups very easily. 
However, in order to make contact with experiments, 
one would have to be able
to predict the value of $\hat{\xi}$.
An upper bound can be obtained from causality,\cite{Kibble:1976sj}
but a more precise estimate requires extra
assumptions about the dynamics of the 
system.\cite{Zurek:1996sj,Rivers:2000mb}

\section{GAUGE INVARIANCE}
If the order parameter field is electrically charged, as it is, e.g.,
in the case of superconductors, 
it couples to the electromagnetic vector potential.
Then the analogue of the approximate free energy~(\ref{equ:Fglobal}) 
is\cite{Kleinert}
\begin{equation}
\label{equ:Flocal}
F[\psi,\vec{A}]=\int d^2x
\left[\frac{1}{2}B^2+\vec{D}\psi^*\cdot \vec{D}\psi
+m^2\psi^*\psi+\lambda\left(\psi^*\psi\right)^2
\right],
\end{equation}
where $B$ is the magnetic induction,\footnote{In three dimensions,
$B$ would, of course, be a vector field $\vec{B}=\vec{\nabla}\times\vec{A}$.}
$B=\partial_xA_y-\partial_yA_x$,
and $\vec{D}$ is the covariant derivative
$\vec{D}=\vec{\nabla}+ie\vec{A}$,
and $\vec{A}$ is the electromagnetic vector potential (gauge field).

The coupling to $\vec{A}$ has the important consequence that the U(1)
symmetry becomes local; the free energy is invariant under all
transformations
\begin{equation}
\psi(\vec{x})\rightarrow\exp(i\Lambda(x))\psi(\vec{x}),\qquad
\vec{A}(\vec{x})\rightarrow\vec{A}(\vec{x})-
\frac{i}{e}\vec{\nabla}\Lambda(\vec{x}),
\end{equation}
where $\Lambda(\vec{x})$ can be any real function of $\vec{x}$.

In particular, this local gauge invariance  
means that the phase angle of $\psi$ does not have
any physical meaning, because it can always be transformed away by a
suitable choice of $\Lambda$. 
For this reason, one cannot talk about domains of correlated $\psi$
anymore. In the global case, the gradient term in the free
energy (\ref{equ:Fglobal}) tends to make the order parameter aligned, 
but in the local case, 
it is the covariant derivative $\vec{D}\psi$ and not the gradient that
will be minimized. In the special case in which $\vec{A}=0$, there is
no difference, but in practice there are always non-zero
thermal fluctuations present in the $\vec{A}$ field. 
If $\vec{A}$ is non-zero, the
covariant derivatives cannot typically be minimized everywhere at the
same time, and this leads to vortex formation.

In fact, this effect can easily be seen in simple cases in which the
initial value of the 
vector potential $\vec{A}$ has been set by hand. As a very simple
example, let us think of a phase transition in the presence of an
homogeneous magnetic field $B(\vec{x})=B_0$. When the system enters
the broken phase, the dynamics tries to minimize $\vec{D}\psi$. Let us
write $\psi(\vec{x})=v(\vec{x})\exp(i\theta(\vec{x}))$, and consider a
circular curve of radius $R$. At every point on this curve, the system
tries to minimize the covariant derivative, but that leads to the result
\begin{equation}
\int_C d\vec{x}\cdot\vec{\nabla}\theta=
\frac{1}{e}\int_C d\vec{x}\cdot\vec{A}=\frac{1}{e}\int_S d^2xB=
\frac{2\pi RB_0}{e}.
\end{equation}
This means that if $RB_0\gsim e$, non-zero winding number is
preferred, and consequently a vortex is formed. This
formation of an Abrikosov vortex lattice is a well-known phenomenon in
superconductor physics.

\section{PLANE WAVE}
In order to understand the dynamics of phase transitions 
in which the magnetic field is inhomogeneous, it is instructive
to consider first a plane wave. Let us assume that there is a standing
electromagnetic wave trapped inside our two-dimensional system,
\begin{equation}
\label{equ:inicond}
B(x,y)=B_0\sin(kx),
\end{equation}
where $k$ is the wave number of the wave. 

The free energy (\ref{equ:Flocal}) by itself determines only the
thermodynamics of the model, not the dynamics.
For that, we need the equations of motion, and here we will consider
a simple, ``relativistic'' model, 
\begin{eqnarray}
\partial^2_0\psi &=& \vec{D}^2\psi-V'(\psi),
\nonumber\\
\partial_0 E_i&=&\epsilon_{ij}\partial_jB+2e{\rm Im}\psi^*D_i\psi,
\nonumber\\
\vec{\nabla}\cdot\vec{E}&=&2e{\rm Im}\psi^*\partial_0\psi,
\label{equ:eom}
\end{eqnarray}
where $V(\psi)=m^2|\psi|^2+\lambda|\psi|^4$.
Note that there is no external magnetic field acting on the system;
instead, Eq.~(\ref{equ:inicond}) merely specifies the initial
conditions
for the time evolution.

\begin{figure}
\centerline{
\includegraphics[height=4cm]{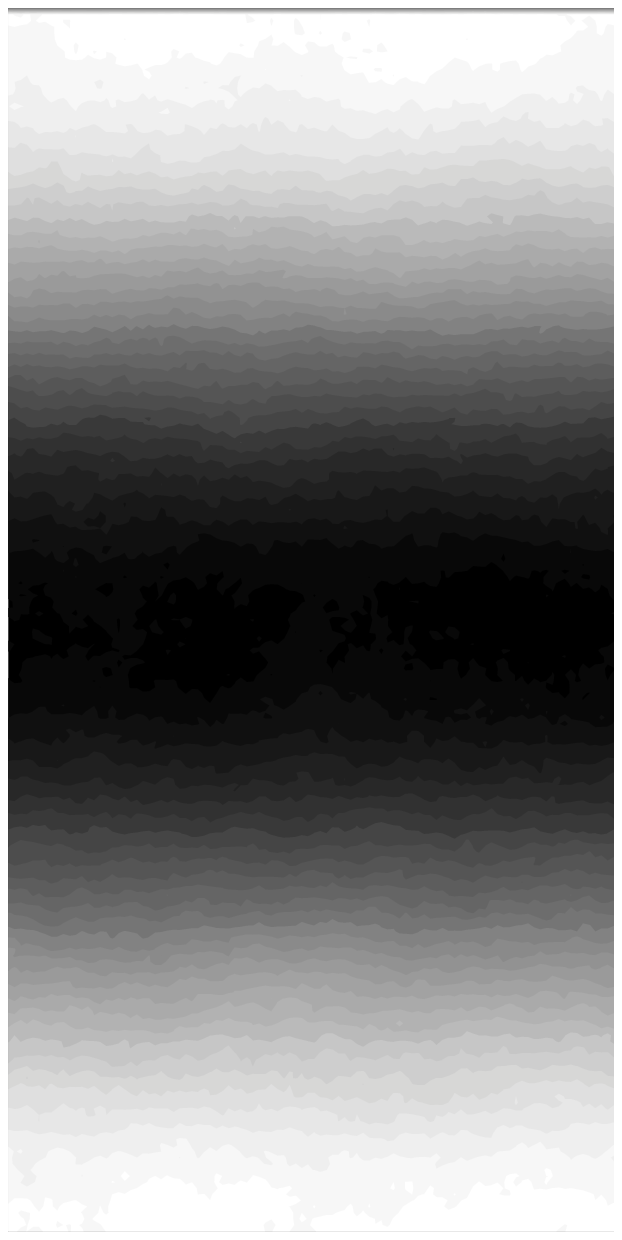}
\includegraphics[height=4cm]{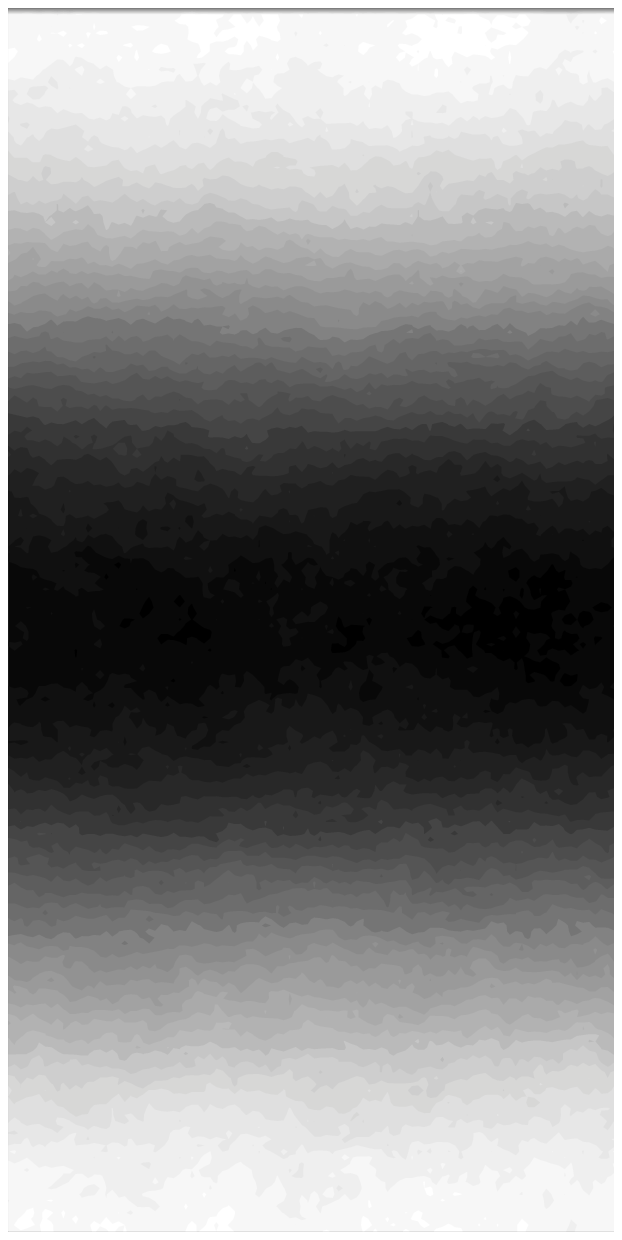}
\includegraphics[height=4cm]{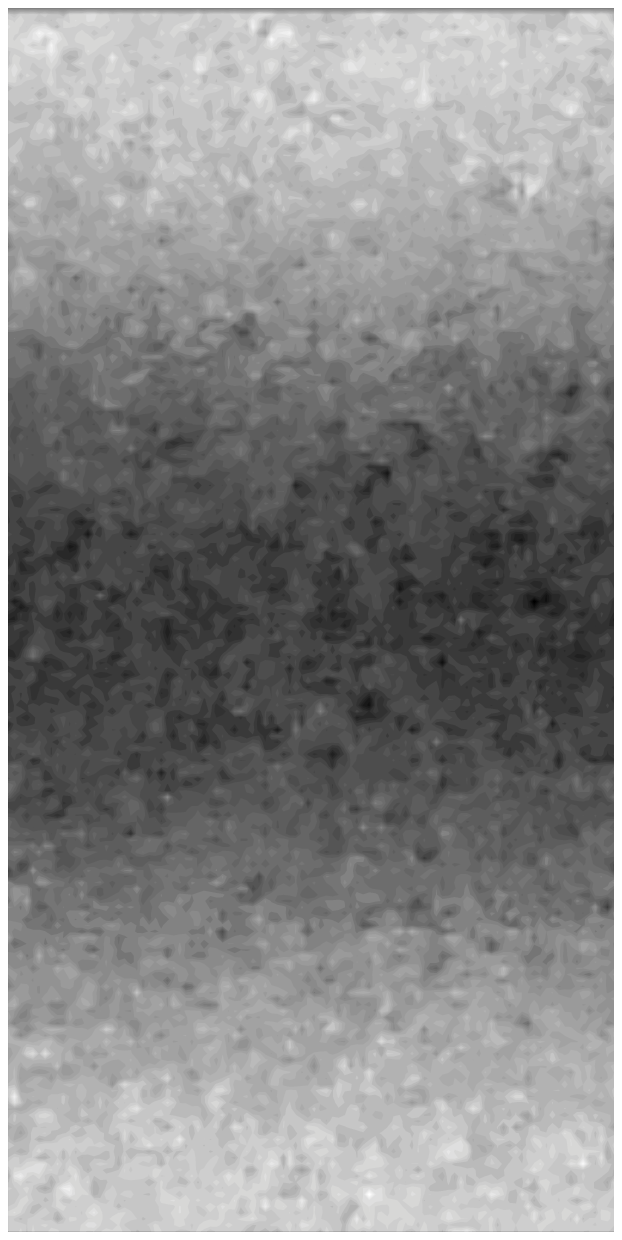}
\includegraphics[height=4cm]{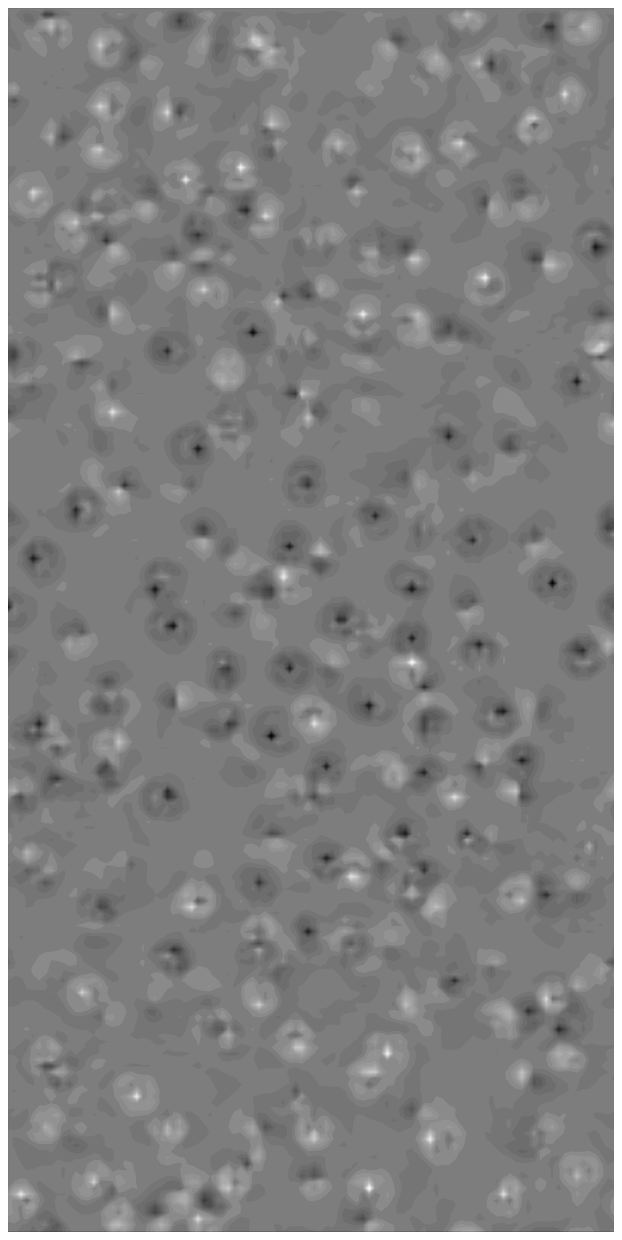}
\includegraphics[height=4cm]{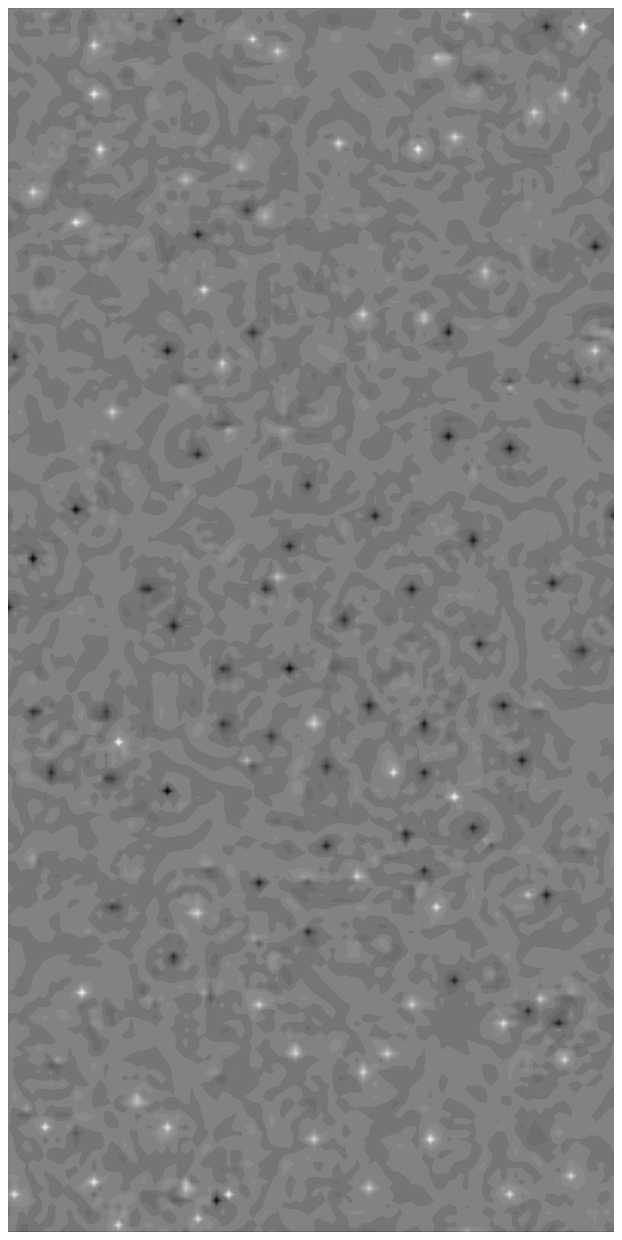}
\includegraphics[height=4cm]{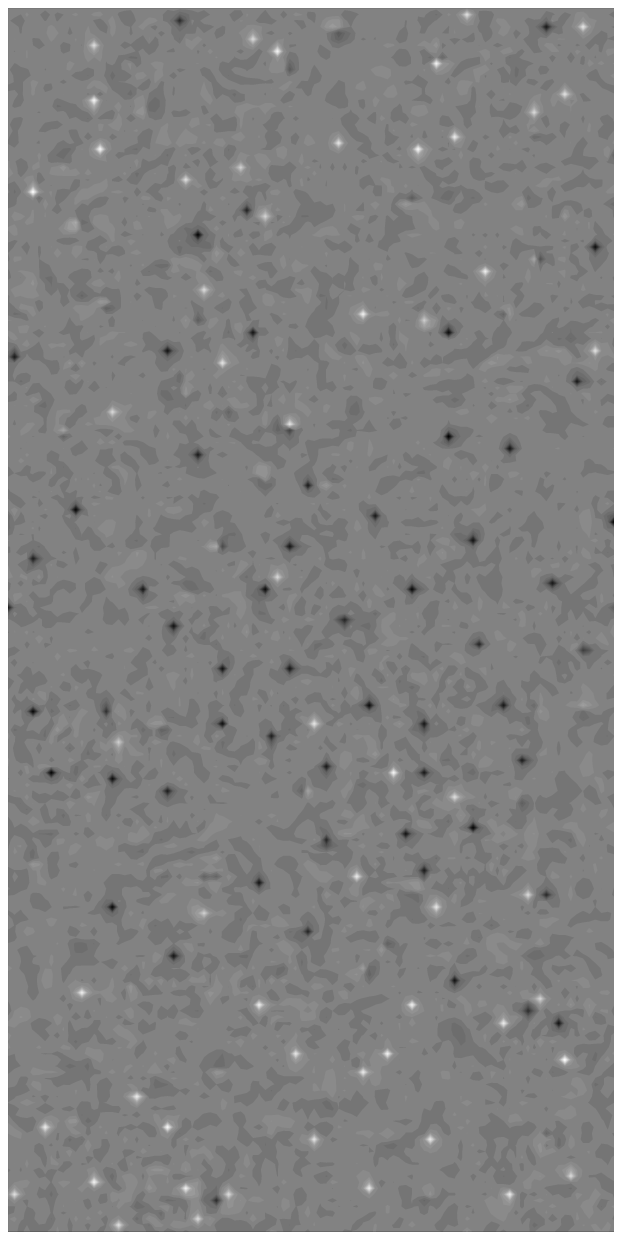}}
\caption{
\label{fig:Bmap}
The time evolution of the local magnetic induction $B(\vec{x})$ in a
slow phase transition, starting from a plane wave configuration.
Shades from white to black indicate field values from positive to negative.
Initially, the amplitude of the field is very small, but the
the contrast of the plot 
is adjusted accordingly to make it visible. In the final
state, the flux is confined into vortices inside which the value of
$B(\vec{x})$ is very high. The signs of the vortices are strongly
correlated with the initial value of $B$ at each point.
}
\end{figure}

Let us now ``quench'' the system into the broken phase by changing
$m^2$ rapidly. To be specific, let us assume that
\begin{equation}
m^2(t)=-m_0^2\frac{t}{\tau_Q}.
\end{equation}
If $\tau_Q$ is small enough in comparison to the the wave number $k$,
the situation in
the regions in which $B$ is, say, positive looks exactly the same as
in the case of a homogeneous magnetic field.
Therefore, vortices with a positive sign will form. Likewise, in the
regions with negative $B$, vortices with a negative sign form.
Fig.~\ref{fig:Bmap} shows the configuration of the $B$ field at different
times during such a transition. In the initial state, the order
parameter field $\psi$ was given tiny random fluctuations, and then
the mass parameter was decreased so that the system underwent a phase
transition to the broken phase. 
The plot shows
clearly how the signs of the vortices are correlated 
with the initial value of
the magnetic induction at each point.
Note, however, that the agreement is not perfect. These vortices with
``wrong'' signs are formed by the Kibble mechanism.

On the other hand, if $\tau_Q$ is large enough, the transition is
adiabatic. The amplitude of $B$ decreases, 
and the system ends up in the equilibrium state, which contains no
vortices because of the absence of any external magnetic fields. 
This shows that for each $k$ there is a critical value of
$\tau_Q$ so that if the transition is faster, vortices will form. We
can also invert this relation and define for each $\tau_Q$ a critical
wavenumber $\hat{k}(\tau_Q)$.

\section{THERMAL EQUILIBRIUM}
In practice, we are not interested in creating vortices by external
magnetic fields or by standing electromagnetic waves. 
Instead, we would like to understand what happens when the transition
starts from thermal equilibrium.
However, similar arguments apply to that case as well, because
thermal fluctuations alone can act as seeds
for vortex formation.

In thermal equilibrium, the state of the system is not described by a
single field configuration, but by an ensemble of configurations in
which the probability of any
given configuration is proportional to $\exp(-E[\psi,\vec{A}]/T)$.
Here $E[\psi,\vec{A}]$ is the energy of the field configuration, and
is essentially equal to the free energy (\ref{equ:Flocal}).
Because of the smallness of the gauge coupling constant $e$, 
we can neglect the coupling to $\psi$ and approximate the energy by
\begin{equation}
E[\vec{A}]\approx\int d^2x \frac{1}{2}B(\vec{x})^2.
\end{equation}
Thus the
probability distribution of $B$ is almost Gaussian, and
the field strength is totally uncorrelated
between all the points in the coordinate space.
If we write the
energy in the Fourier space, we find that the same is true there
as well,
\begin{equation}
E[\vec{A}]\approx\int\frac{d^2k}{(2\pi)^2}\frac{1}{2}|B(\vec{k})|^2,
\end{equation}
i.e., the amplitude of each Fourier mode independent of all the others
and has a Gaussian distribution with a $k$-independent width.
As long as the system is in the symmetric phase, each mode also behaves
dynamically independently of the other modes. Therefore we can
estimate that in a transition with quench time $\tau_Q$, 
the modes with $|\vec{k}|\lsim\hat{k}(\tau_Q)$ equilibrate, but the
modes with $|\vec{k}|\gsim\hat{k}(\tau_Q)$ fall out of equilibrium and
their distribution remains the same as it was in the symmetric phase
before the transition. 

To make comparison with the Kibble mechanism easier, we denote the
wavelength corresponding to $\hat{k}$ by $\hat{\xi}=2\pi/\hat{k}$.
The above implies that if we concentrate on a region $S$ of radius
$\hat{\xi}$, the total magnetic flux through this region
does not change significantly in the transition. 
Inside the region, the flux rearranges itself in order to minimize the
energy, i.e., it forms a vortex ``lattice'' in which all the vortices have
equal sign.
Therefore, we can
estimate the number of vortices in the region $S$ after the
transition by calculating the typical flux through it at the time of
the transition, and this is approximately the same as it is in the symmetric
phase. Because of the Gaussianity of the probability distribution,
it is easy to calculate,
\begin{equation}
\left(\Phi_S^{\rm typ}\right)^2\approx
\langle\Phi_S^2\rangle
=\int d^2x d^2x'\langle B(\vec{x})B(\vec{x}')\rangle
=\int d^2x T=\pi\hat{\xi}^2T.
\end{equation}
Therefore, the typical number of vortices through the region $S$ in the
broken phase is
\begin{equation}
N_S=\Phi_S^{\rm typ}/\Phi_0\approx
\frac{e}{2\pi}T^{1/2}\hat{\xi},
\end{equation}
and, consequently, the number density of vortices is
\begin{equation}
\label{equ:2dpred}
n\approx \frac{N_S}{\pi\hat{\xi}^2} \approx
\frac{e}{2\pi}T^{1/2}\hat{\xi}^{-1}. 
\end{equation}

Note that this estimate was for a two-dimensional system, and
therefore $e$ is not a dimensionless number like it is in three
dimensions, but instead, it has a dimension of energy$^{1/2}$.
The three-di\-men\-sion\-al case has also some other differences, and it
will be discussed in Section~\ref{sect:3d}

The essential point in Eq.~(\ref{equ:2dpred}) is the relation
$n\propto\hat{\xi}^{-1}$, which differs from the prediction of the
Kibble mechanism $n\propto\hat{\xi}^{-2}$ in Eq.~(\ref{equ:kibblepred}).
In order to really predict the vortex number density in either case,
we must be able to calculate $\hat{\xi}$ and that depends sensitively
on the dynamics of the system. However, if we assume that the critical
wavelength $\hat{\xi}$ is or the same order of magnitude in both cases,
Eq.~(\ref{equ:2dpred}) becomes greater than Eq.~(\ref{equ:kibblepred})
when $\hat{\xi}\gsim 2\pi/eT^{1/2}$.
Because long wavelengths correspond to slow transitions, we can
conclude that this mechanism dominates in slow transitions.

\section{SPATIAL DISTRIBUTION}
\begin{figure}
\centerline{\includegraphics[height=4cm]{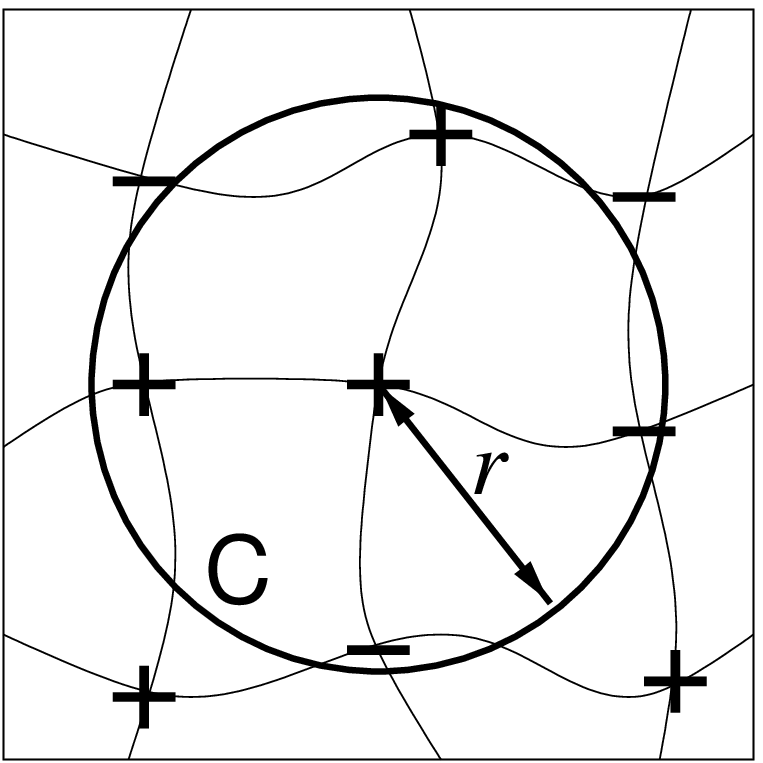}
\includegraphics[height=4cm]{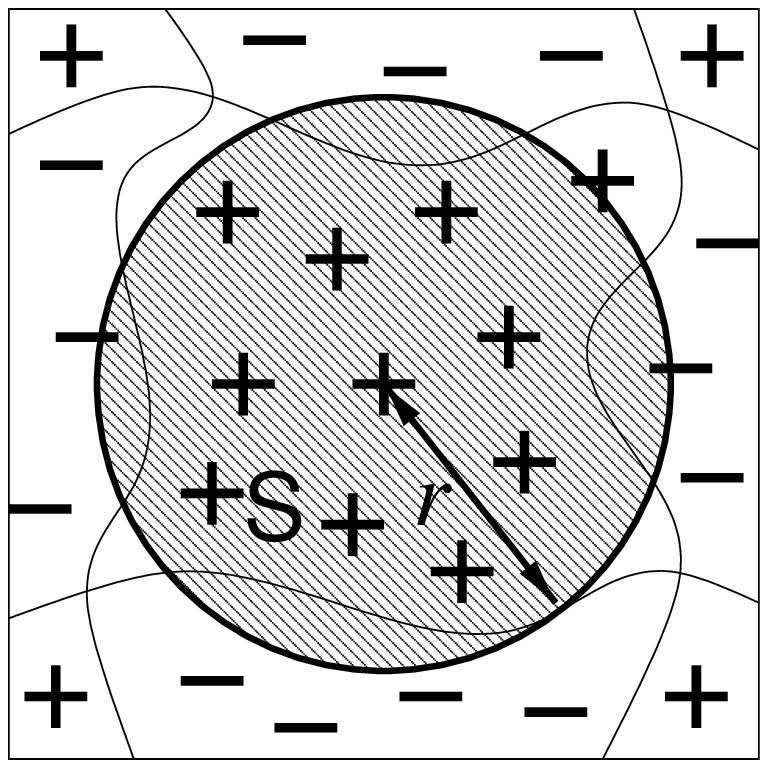}}
\caption{
\label{fig:Ncr}
The winding number of a circular curve of radius $r$
centred at a positive vortex, $N_C(r)$, behaves differently in the
Kibble mechanism (left) and in the local gauge theory (right).
}
\end{figure}

Even without any assumption on $\hat{\xi}$,
our mechanism can be tested
by studying the spatial distribution of vortices. One particularly
simple way of seeing this is by considering the average winding number
along a circular curve of radius $r$ centred at a vortex of positive
sign
(see Fig.~\ref{fig:Ncr}). In
practice, we pick a vortex, draw a circle of radius $r$ around it
and count the number of vortices of the same sign as the one at the
centre minus the number of vortices of opposite sign, and average this
over all vortices. We denote this quantity by $N_C(r)$. Obviously,
$N_C(0)=1$, because then only the vortex itself is counted.

Let us first consider the Kibble mechanism.
In this case, the vortex number is determined by the phase angle
$\gamma=\arg\psi$ of the order parameter field, and we can write
\begin{equation}
\label{equ:NCrkibble}
N_C(r)=\int_C d\vec{r}\cdot\vec{\nabla}\gamma.
\end{equation}
The vortices are formed at the points where domains meet, and
therefore the distance to the nearest vortex is roughly the domain
size $\hat{\xi}$. Therefore, as long as $r\ll\hat{\xi}$, $N_C(r)$ does
not change significantly as we increase $r$.

However, when $r\gg\hat{\xi}$, every point on the curve $C$ is
uncorrelated with the centre of the curve, because the correlation
length $\hat{\xi}$ is less than $r$. Therefore the integral
(\ref{equ:NCrkibble}) must be independent of whether there is a
positive or a negative vortex at the centre. This implies that it
cannot have a preferred sign, and therefore it must
vanish on average. Summarizing, we find that in the Kibble mechanism
\begin{equation}
\label{equ:NCrkibblepred}
N_C(r)\approx\left\{
\begin{array}{ll}
1,& r\lsim \hat{\xi},\cr
0,& r\gsim \hat{\xi}.
\end{array}
\right.
\end{equation}

In our mechanism, the vortex number is determined by the magnetic
field, and therefore it is more convenient to write\footnote{
In a suitably chosen gauge, Eq.~(\ref{equ:NCrkibble}) is valid in the
gauge theory as well, but the arguments that lead to the
prediction~(\ref{equ:NCrkibblepred}) are not.
}
\begin{equation}
\label{equ:NCrgauge}
N_C(r)=\int_{S(C)} d^2x B(\vec{x}),
\end{equation}
where $S(C)$ is the region bounded by the curve $C$. In this case, the
vortices are inside domains and all the vortices in a given domain
have equal sign. When we start increasing $r$ from zero, each new
vortex we pick up has the same sign as the one at the origin, as long
as $r\ll\hat{\xi}$. This implies that
$N_C(r)$ increases with increasing
$r$. However, when $r\gg\hat{\xi}$, the sign of any new vortex
encountered is independent of the vortex at the
centre, and therefore, on average, $N_C(r)$ remains constant. Thus
we conclude that
\begin{equation}
\label{equ:NCrgaugepred}
N_C(r)\approx\left\{
\begin{array}{ll}
1+cr^2,& r\lsim \hat{\xi},\cr
{\rm constant}~(\ge 1),& r\gsim \hat{\xi},
\end{array}
\right.
\end{equation}
where $c$ is some constant.

The two predictions (\ref{equ:NCrkibblepred}) and
(\ref{equ:NCrgaugepred}) are qualitatively different, and therefore it is easy
to determine which of the mechanisms dominates if one is able to
measure the
spatial distribution of vortices in the final state. In 
numerical simulations, this is straightforward, and we have carried out
this measurement.\cite{Hindmarsh:2000kd}

In our simulations, we used a thin (120$\times$120$\times$5)
three-di\-men\-sion\-al
lattice, and
solved numerically the the equations of motion
\begin{eqnarray}
\partial^2_0\psi &=& \vec{D}^2\psi-V'(\psi),
\nonumber\\
\partial_0 \vec{E}&=&\vec{\nabla}\times\vec{B}+2e{\rm Im}\psi^*\vec{D}\psi,
\nonumber\\
\vec{\nabla}\cdot\vec{E}&=&2e{\rm Im}\psi^*\partial_0\psi.
\label{equ:eom3d}
\end{eqnarray}
The time step was $\delta t=0.05$ in units where the lattice spacing
is one.
The
coupling constants were $e=0.3$ and $\lambda=0.18$. Because
$\lambda>e^2$, this corresponds to a type-II superconductor.
We first thermalized the system to temperature $T=6$ in the symmetric
phase using a hybrid Monte Carlo algorithm. Then, we evolved the
system according to the equations of motion (\ref{equ:eom3d}), changing
the mass term as
\begin{equation}
m^2(t)=m_0^2-\delta m^2\left(\frac{4}{3\pi}\arctan\frac{t}{\tau_Q}+\frac{1}{3}
\right).
\label{equ:masschange}
\end{equation}
\begin{figure}
\centerline{\includegraphics[height=8cm]{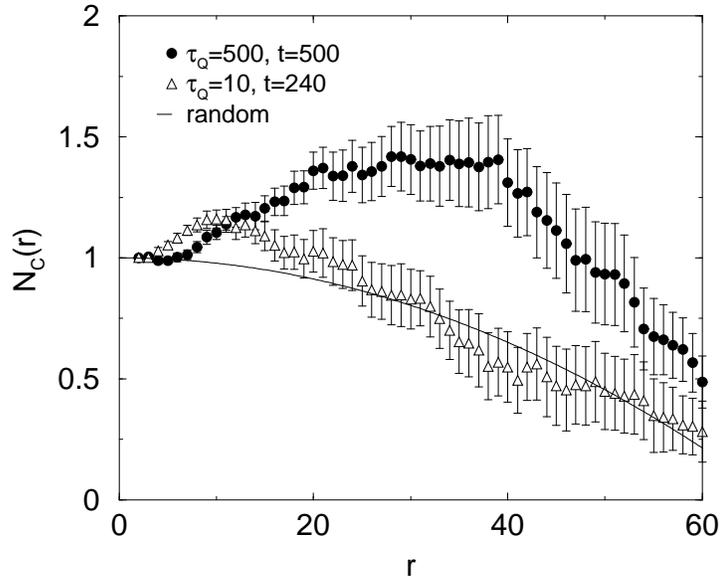}}
\caption{
\label{fig:wcorr}
The average of $N_C(r)$ as a function of $r$ after slow
($\tau_Q=500$) and fast ($\tau_Q=10$) 
transitions (from Ref.~\onlinecite{Hindmarsh:2000kd}).
The solid line is a
benchmark curve, which shows the behaviour in the case where there are
no correlations at all between vortices. In both cases the data
points are significantly above this curve at short distances, which
indicates positive correlations at short distances, just like
predicted by Eq.~(\ref{equ:NCrgaugepred}). The signal gets stronger in
slow transitions.}
\end{figure}%
Here $m_0^2=1.4$ is the initial value at 
$t=-\tau_Q$ when the quench begins and
$\delta m^2=3.2$ was chosen
in such a way that the transition takes place at $t\approx 0$.
In the final state, we measured the function $N_C(r)$ and averaged it
over several initial configurations taken from the thermal ensemble.
The results are shown in Fig.~\ref{fig:wcorr} for two different quench
rates. The solid line in the plot is to illustrate the effects of the
finite system size. Because the total flux through the system is fixed
to zero by the boundary conditions, $N_C(r)$ would not be unity even
for a random distribution of vortices. Instead, it would decrease as
$N_C(r)=1-\pi r^2/A$, where $A=120^2$ is the cross-sectional area of
the lattice. This is shown as the solid curve in Fig.~\ref{fig:wcorr}.
The fact that the data points are significantly above it
indicates that the vortex distribution indeed behaves as in
Eq.~(\ref{equ:NCrgaugepred}), supporting our mechanism.

\section{THREE DIMENSIONS}
\label{sect:3d}
A superconductor film resembles in many respects the thin
three-di\-men\-sion\-al model discussed in the previous section, and
therefore it might seem that the result would apply directly to a
real experimental situation. However, 
there is an important difference: In reality the magnetic field lives
in three dimensions, and the part of it that is outside the
superconductor also contributes to the free energy.

For simplicity, we shall now assume that in the symmetric phase, the
magnetic field behaves as if it was in vacuum. This is not a totally
unreasonable approximation, because if the superconducting film is
thin, the energy of a typical field configuration is dominated by the
contribution from the vacuum outside the film. 

Furthermore, we assume that when the film becomes superconducting, it
starts to dominate the dynamics of the magnetic field. Essentially
this means that 
we can treat the vortices, instead of the magnetic field, as the
relevant degrees of freedom.

Again, we have to estimate the typical flux through a circular loop of
radius $\hat{\xi}$ on the film. The configuration of minimal energy
in which the flux is $\Phi$, is simply the magnetic dipole, and its
energy is roughly
\begin{equation}
E(\Phi,\hat{\xi})\sim\frac{\Phi^2}{\mu_0\hat{\xi}}.
\end{equation}
We can estimate that thermal fluctuations can generate this
configuration if its energy is less than the temperature, $E\lsim
k_BT$. 
This
reasoning leads to
\begin{equation}
\Phi\sim \sqrt{\mu_0k_BT\hat{\xi}}=\frac{e}{\pi\hbar}\left(
\mu_0k_BT\hat{\xi}\right)^{1/2}
\times\Phi_0.
\end{equation}
Dividing this by the area of the loop, we find the number density
\begin{equation}
n\sim\frac{e}{\pi\hbar}\left(
\frac{\mu_0k_BT}{\hat{\xi}^3}\right)^{1/2}.
\end{equation}

It is more difficult to estimate $\hat{\xi}$, because it depends on 
the dynamical properties of the superconductor near the transition
point,
and we don't attempt to do that here.
Instead, we calculate how large $\hat{\xi}$ must be in a realistic
experiment so that the correlations between vortices would be clear
enough to be observed. This requires that the number of vortices
inside each domain is much greater than one, i.e.,
\begin{equation}
N_S\approx 
\frac{e}{\pi\hbar}\left(
\mu_0k_BT\hat{\xi}\right)^{1/2}\gg1.
\end{equation}
The critical temperature of, e.g., YBCO superconductors is $T=90$~K,
and this leads to the estimate
\begin{equation}
\hat{\xi}\gg\frac{\pi^2\hbar^2}{e^2\mu_0k_BT}\approx 3~{\rm mm},
\end{equation}
which is a very reasonable size from the experimental point of view.

Experiments like this have not been carried out yet, and there may
well be technical difficulties involved in the measurement of the
spatial distribution of the vortices after the transition. In
particular, one will have to be able to freeze the locations of the
vortices so that they don't annihilate before the measurement has been
finished. 

On the other hand, an otherwise very similar experiment has
been carried out recently by Carmi et al.\cite{ref:carmi} Instead of
measuring the number and locations of individual vortices, they
measured the total net flux, i.e., the number of positive vortices
minus the number of negative vortices. The result they found was that
the flux was consistent with zero, which contradicts the predictions
of the Kibble mechanism. The mechanism presented in this paper cannot
explain this result, because it would typically predict an extra set
of vortices in addition to those formed by the Kibble
mechanism.
It does not make the discrepancy with the experiment any worse
either, because
our mechanism does not actually change the total flux through the film
at all, and therefore the variation in the total net flux in the final
state is simply equal to the variation in the initial flux, i.e., very
small.

\section{CONCLUSIONS}
I have discussed the scenario of defect formation 
presented in Ref.~\onlinecite{Hindmarsh:2000kd} 
from the point of view of superconductors. This
scenario is not merely a modification to the standard Kibble
mechanism, but a totally separate mechanism, and in a general
phase transition, both of them form vortices. Which one of them
dominates, depends on the details of the transition.

Although estimating the density of vortices formed in the transition
by either mechanism
requires a detailed knowledge about the dynamics of the system, it is
possible to derive simple and robust predictions for the spatial
distribution of vortices. Basically, the Kibble mechanism predicts
negative correlations between vortices, i.e., that nearby vortices
have opposite signs, whereas the mechanism discussed here predicts
positive correlations. These predictions can be tested experimentally,
if it is possible to measure the locations and signs of individual
vortices after the phase transition. The domain
size needed for these vortex-vortex correlations to be significant in YBCO
superconductors, 
being of the
order of centimeter, is suitable for experiments.

\section*{ACKNOWLEDGMENTS}
I would like to thank Mark Hindmarsh for collaboration on this
subject, and Ray Rivers, Erkki Thuneberg and Grisha Volovik for useful
discussions.
The author was supported by PPARC and also partly by the University 
of Helsinki.
Part of this work was conducted on the SGI Origin platform using COSMOS
Consortium facilities, funded by HEFCE, PPARC and SGI.

\end{document}